\documentclass[iop,numberedappendix,appendixfloats]{emulateapj}

\usepackage{epstopdf}
\usepackage{graphics}
\usepackage{natbib}
\usepackage{hyperref}

\newcommand{\kms}{\mbox{km s$^{-1}$}}

\newcommand{\etal}[1]{{ et al.}~}
\def\kms{\ifmmode \hbox{km~s}^{-1}\else km~s$^{-1}$\fi}
\def\etal {{\it et al.}}
\def\deg      {{\ifmmode^\circ\else$^\circ$\fi} } 

\def\h2     {H$_2$}

\shorttitle{High Redshift Dust Attenuation}
\shortauthors{Scoville \etal}

\begin{document}

\title{Dust Attenuation in High Redshift Galaxies -- \\ 'Diamonds in the Sky'}
 \author{Nick Scoville\altaffilmark{1}, Andreas Faisst\altaffilmark{3}, Peter Capak\altaffilmark{1,2},  Yuko Kakazu\altaffilmark{4}, Gongjie Li\altaffilmark{5}, and Charles Steinhardt\altaffilmark{1,2}}
 
\altaffiltext{1}{California Institute of Technology, MC 249-17, 1200 East California Boulevard, Pasadena, CA 91125}
\altaffiltext{2}{Infrared Processing and Analysis Center (IPAC), 1200 E. California Blvd., Pasadena, CA, 91125, USA}
\altaffiltext{3}{Institute for Astronomy, Swiss Federal Institute of Technology (ETH Zurich), CH-8093 Zurich, Switzerland}
\altaffiltext{4}{Subaru Telescope, 650 N. A`ohoku Place, Hilo, HI 96720}
\altaffiltext{5}{Harvard-Smithsonian Center for Astrophysics, The Institute for Theory and Computation, 60 Garden Street, Cambridge, MA 02138}

\altaffiltext{}{}

\begin{abstract}
We use observed optical to near infrared spectral energy distributions (SEDs) of 266 galaxies in the COSMOS 
survey to derive the wavelength dependence of the dust attenuation at high redshift. All of the galaxies 
have spectroscopic redshifts in the range z = 2 to 6.5. The presence of the CIV absorption 
feature, indicating that the rest-frame UV-optical SED is dominated by OB stars, is used to select objects 
for which the intrinsic, unattenuated spectrum has a well-established shape. Comparison of this 
intrinsic spectrum with the observed broadband photometric SED then permits derivation of the wavelength dependence of the
dust attenuation. The derived dust attenuation curve is similar in overall shape to the Calzetti curve for local 
starburst galaxies. We also see the 2175 \AA~bump feature which is present in the Milky Way and LMC 
extinction curves but not seen in the Calzetti curve. The bump feature is commonly attributed to graphite or PAHs. No significant dependence is seen with redshift between sub-samples 
at z = 2 - 4 and z = 4 - 6.5. The 'extinction' curve obtained here provides a firm 
basis for color and extinction corrections of high redshift galaxy photometry. 

\end{abstract}

\medskip 


\section{Introduction}

Dust profoundly affects the light emitted from galaxies -- requiring large corrections for extinction in the UV and optical.
This causes a severe degeneracy in the derived ages of  high redshift galaxies  
and hence results in a major uncertainty in their derived properties. The extinction has a major effect on the observability of the galaxies and on their derived stellar masses, star formation rates and luminosities. At early epochs there is an increasing fraction 
of active star forming galaxies compared to the low redshift local galaxy population. The observed properties of these star forming galaxies are strongly dependent on the dust extinction, since many of the surveys probe the rest frame UV,
where the dust extinction is higher than at visible wavelengths. The amount of dust extinction is almost always inferred 
from the differential extinction or reddening. Translating this reddening  
into an estimate of the total extinction at any wavelength then requires knowledge of the dust extinction curve as 
a function of wavelength.

\begin{figure*}[ht]
\epsscale{0.8}
\plotone{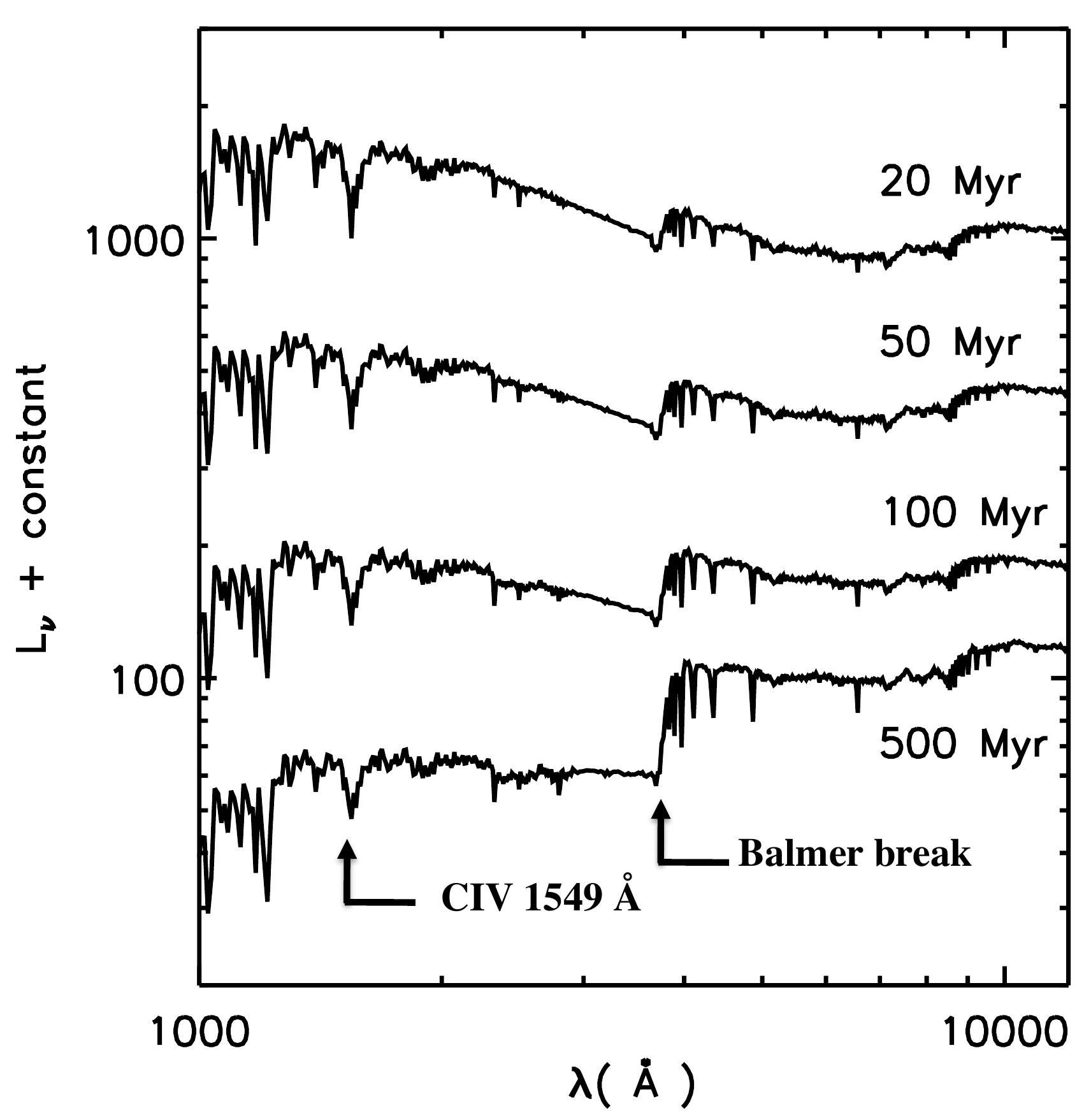}
\caption{Simulated spectra for \emph{constant} star formation rate (SFR) starbursts of duration 20 to 500 Myr obtained 
using Starburst99 with a Kroupa IMF and solar metallicity. The strong CIV absorption at $\lambda = 1549 \AA$ seen in these spectra requires significant SF within the 
previous 10 Myr since this feature is produced in O-type stars. The overall shape of the SED is relatively invariant for the first $\sim50$ Myr; after that time, the Balmer break at $\lambda = 3646$ \AA becomes more pronounced
and thus causes the broadband SED to change in shape. Any objects with a Balmer break larger than 25\%, corresponding to spectra with ages exceeding that of the model shown with 100 Myr age were removed from our sample. Here we use the combination -- strong CIV absorption 
plus the absence of a strong Balmer/4000\AA~break -- to select sources to illuminate the dust attenuation.
  \label{spectra}}
\end{figure*}

At low redshift the extinction curve has been determined along many lines-of-sight in the Galaxy \citep[cf.][]{sav79}
and in the lower metallicity Large Magellanic Cloud (LMC) and Small Magellanic Cloud (SMC) \citep{pre84,fit86}. The mean extinction curves for each of these 
galaxies are quite different: the SMC curve has the steepest power-law form and no 2175\AA~bump, whereas the LMC and the Milky Way extinction curves both show the 2175\AA~feature and the Milky Way has the flattest overall wavelength dependence \citep{gal10}. 
The SMC has the lowest metallicity of the three galaxies and this probably results in a different grain composition and/or  size distribution. These large differences underscore the critical importance of empirical determination of the reddening curves at high redshifts. An effective extinction curve for the integrated light from galaxies  was obtained by \cite{cal94,cal96} and this is commonly used for analysis of high redshift galaxies. 

Direct determination of the differential extinction curve in high-z galaxies has been difficult using just photometry since one needs to break the age/reddening degeneracy in the observed galaxies and this can't be done without prior assumption of a form for the extinction curve. \cite{kri13} fit a grid of SED templates and power-law attenuation curves to the photometry of a sample of galaxies at z = 0.5 to 2. They used the broad- and intermediate-band photometry to classify the intrinsic 
SED type of each galaxy and then correlated the extinction curve slope and the 2175\AA~bump strength with galaxy type. A similar procedure, fitting a library of SED templates,  was employed by \cite{bua13} to analyze a sample of galaxies at z  = 0.95 - 2.2. Their favored attenuation curve was similar to 
that in the LMC super-shell region. 

The approach we advocate here relies upon the intrinsic or un-extincted spectrum of the stellar population being known or identified  
 {\it a priori} from high resolution spectroscopic features. Then, the differential extinction can be easily derived by comparison of the observed spectral energy distributions (obtained from broadband photometry) with that of the intrinsic spectrum.  In fact, there is a UV spectral signature which clearly identifies the nature of the intrinsic spectrum -- this is  
the CIV absorption feature at $\lambda = 1549$ \AA. The CIV absorption indicates the presence of type O and supergiant B stars and when the CIV absorption is strong, such stars are dominating the UV continuum emission.

The CIV absorption feature in spectra of high redshift galaxies in the COSMOS survey thus selects
objects for which the broadband photometry can be used to determine the differential extinction curve. For such galaxies, 
the CIV absorption implies that the UV-optical spectral energy distribution (SED) is dominated by OB stars and the overall 
slope of the intrinsic (un-extincted) continuum SED is known. 

We use  a sample of 266 galaxies at z  = 2 to 6.5 for which rest-frame UV 
spectroscopy shows the CIV absorption to analyze the wavelength dependence of the dust attenuation.  Starburst (SB) model simulations in \S \ref{theory} define the age range (up to   $\sim$50 Myr) of stellar populations for which the CIV absorption can be used as a signpost for the overall SED. In \S \ref{sample}
we describe the galaxy sample used for this study and then outline the numerical technique in \S \ref{technique}. The derived attenuation curve is presented 
in \S \ref{results} and discussed in \S \ref{discussion}.

\section{CIV Absorption as a Signpost for the SED}\label{theory}

In Figure \ref{spectra} spectra of starburst stellar populations are shown for \emph{constant} star formation rates 
extending from 20 to 500 Myr. These spectra were generated using the Starburst99 simulation tool \citep{lei99}
with solar metallicity and a Kroupa initial mass function for the stars. The strongest spectral feature in these plots
is the CIV absorption at $\lambda = 1549 \AA$~-- this absorption is produced in the atmospheres and outflow winds of O stars, and is present as long as O-type stars dominate the rest frame UV emission. In an instantaneous starburst, the feature 
will diminish rapidly since the typical 
lifetime of those stars is $\leq 10$ Myr. Hence, this feature is a robust signpost of on-going OB star formation. Then, the 
UV-optical continuum, dominated by OB stars, has a quite constant SED shape. 

At longer wavelengths, the dominant spectral feature is the Balmer break at $\lambda = 3646$ \AA. As the 
stellar populations ages, or as the fraction of intermediate mass stars grows, this feature
becomes more pronounced, causing the optical SED shape to evolve -- specifically becoming redder 
at later epochs. 

At early epochs ($\lesssim$20 - 100 Myr) the spectra are very constant in both broadband shape and spectral features (see Figure \ref{spectra})  and such sources 
may be used as a background source to probe the wavelength dependence of the foreground dust 
attenuation. The signatures of appropriate galaxies to select are: 1) the detection of CIV absorption and 2) 
the absence of a strong Balmer/4000\AA~break feature. The CIV absorption feature at $\lambda = 1549 \AA$~ is also seen in type 2 AGN sources, but these can be recognized using AGN signatures such as X-ray and strong radio emission or nuclear point sources. These discriminants against AGN may not catch all AGN at the highest redshifts given the limited sensitivity 
of X-ray and radio data. The CIV absorption is very strong and hence easily detected. For a Kroupa 
IMF, the CIV equivalent width is $\sim 9.5$ \AA~at early times; it can therefore be easily seen down to 1/10 solar metallicity. This 
was confirmed by additional starburst simulations not shown here.

\section{Galaxy Sample Selection}\label{sample}

 \begin{figure}[ht]  
\epsscale{1.}
\plotone{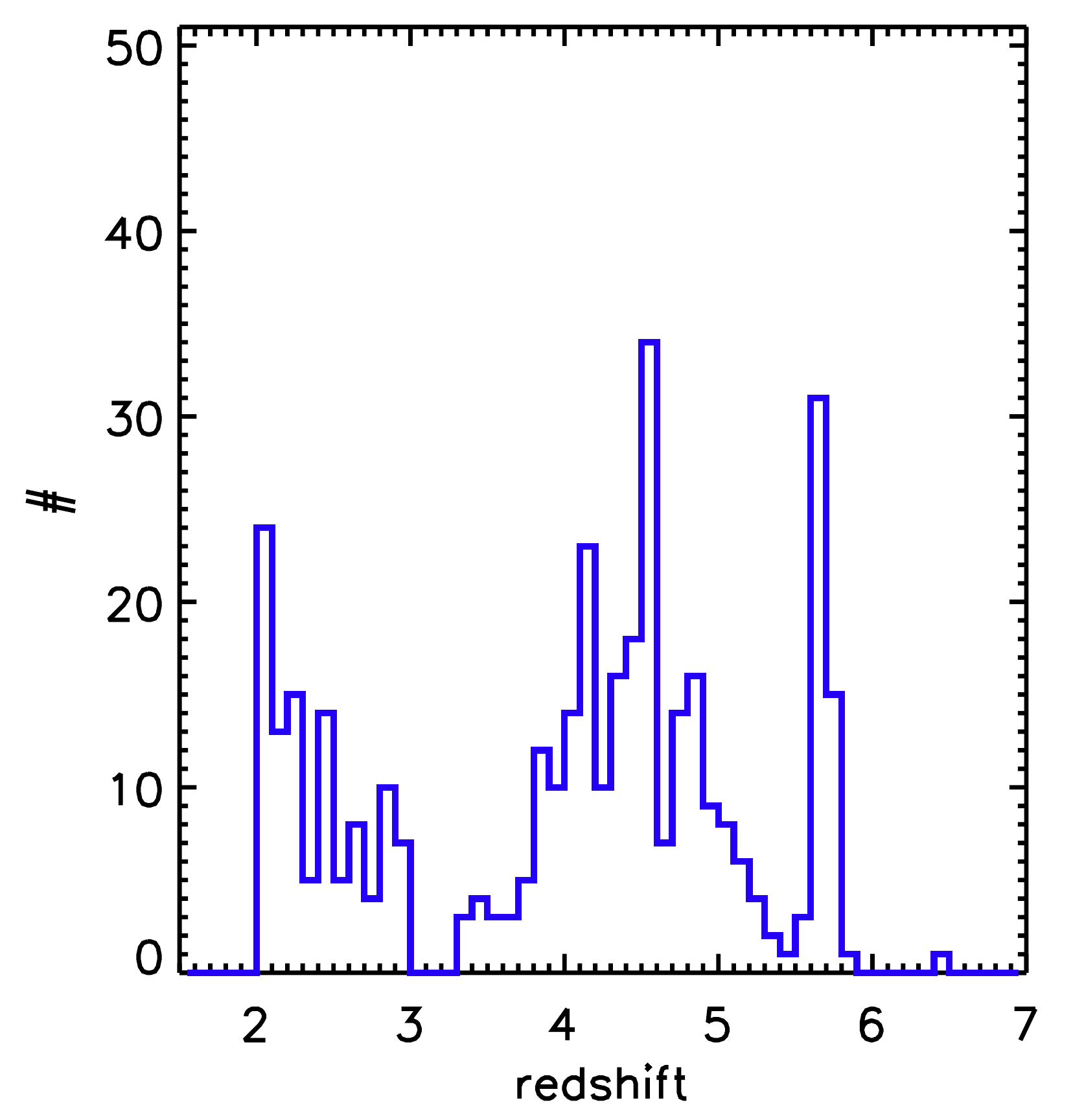}
\caption{The spectroscopic redshift distribution of galaxies in our sample.  \label{redshift_plot}}
\end{figure}

 \begin{figure}[ht]  
\epsscale{1.}
\plotone{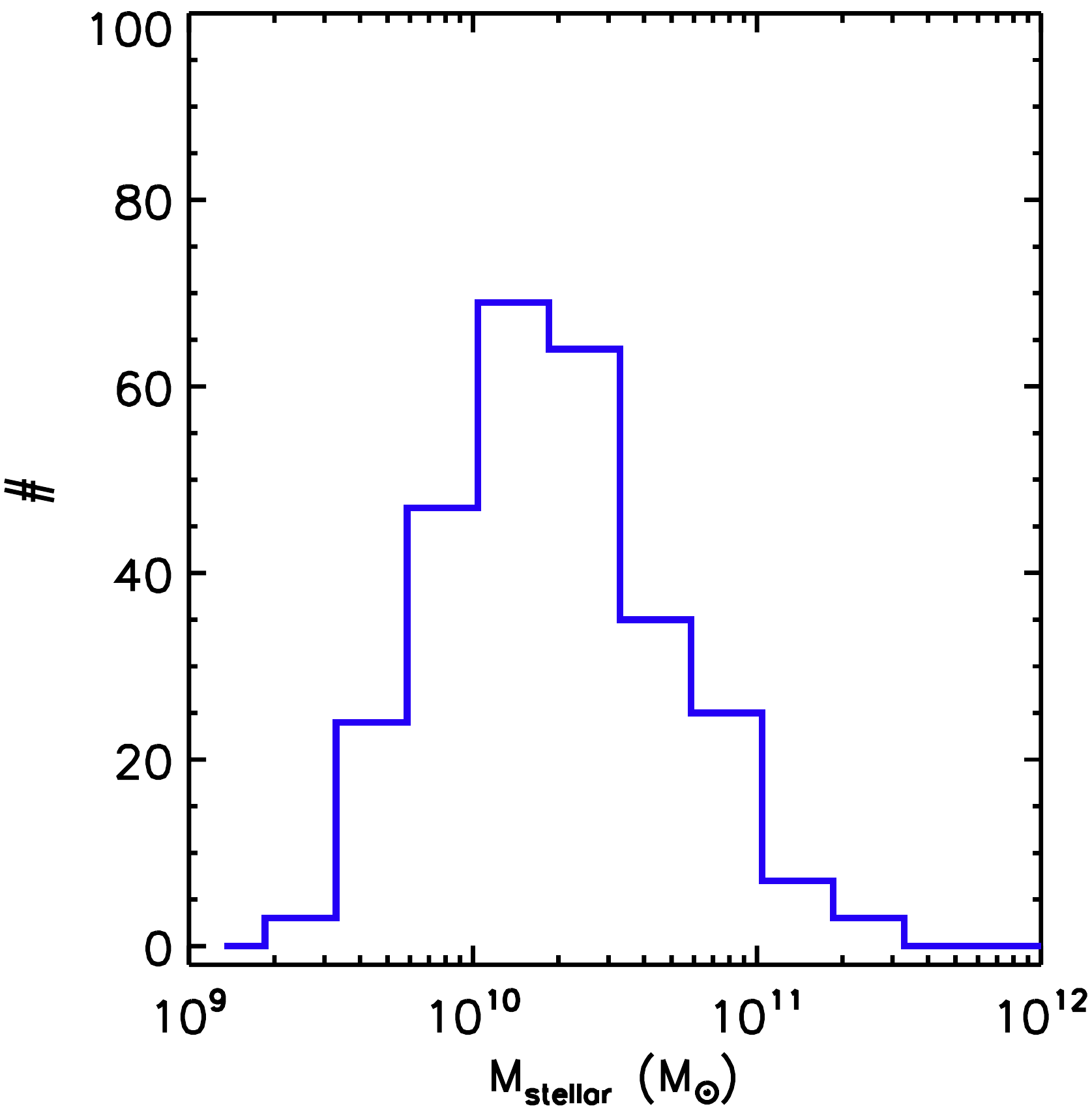}
\caption{The stellar mass distribution of galaxies in our sample. The masses 
were derived from SED template fitting to the photometry using the  ZEBRA+ code with a constant SFH, a Chabrier IMF (similar to Kroupa)  and a metallicity range from Z=0.005, 0.001, 0.01. \label{mass_plot}}
\end{figure}

Our source sample is drawn from the COSMOS  survey field, with the primary selection being those 
objects with spectroscopic coverage from our Keck DEIMOS survey (Scoville, PI). This survey includes $\sim 2300$ 
galaxies at z  = 2 to 6.5 down to I$_{AB} = 25$ mag. From this sample, we then selected only those
objects with the CIV absorption feature and at least 6 broadband continuum photometric detections (average is 10.3 photometric 
bands for each source). The redshifts 
were determined from the CIV absorption plus other emission lines and we required at least two lines for a reliable
spectroscopic redshift. As noted in \S \ref{theory}, we then removed from further consideration any objects exhibiting 
a Balmer break larger than 25\% in flux. The final fitting included 266 objects at z = 2 to 6.5 and then subsamples of 
 135 and 132 objects at z = 2 - 4 and 4 - 6.5.\footnote{The source list is available in an ASCII table from the COSMOS data archive 
 at IPAC/IRSA: \\ \url{http://irsa.ipac.caltech.edu/data/COSMOS/tables/extinction/extinction_curve_source_list.dat}.} The redshift and stellar mass distributions of sources are shown in Figures \ref{redshift_plot} and  \ref{mass_plot}. 
 
\begin{figure*}[ht]
\epsscale{1.}
\plotone{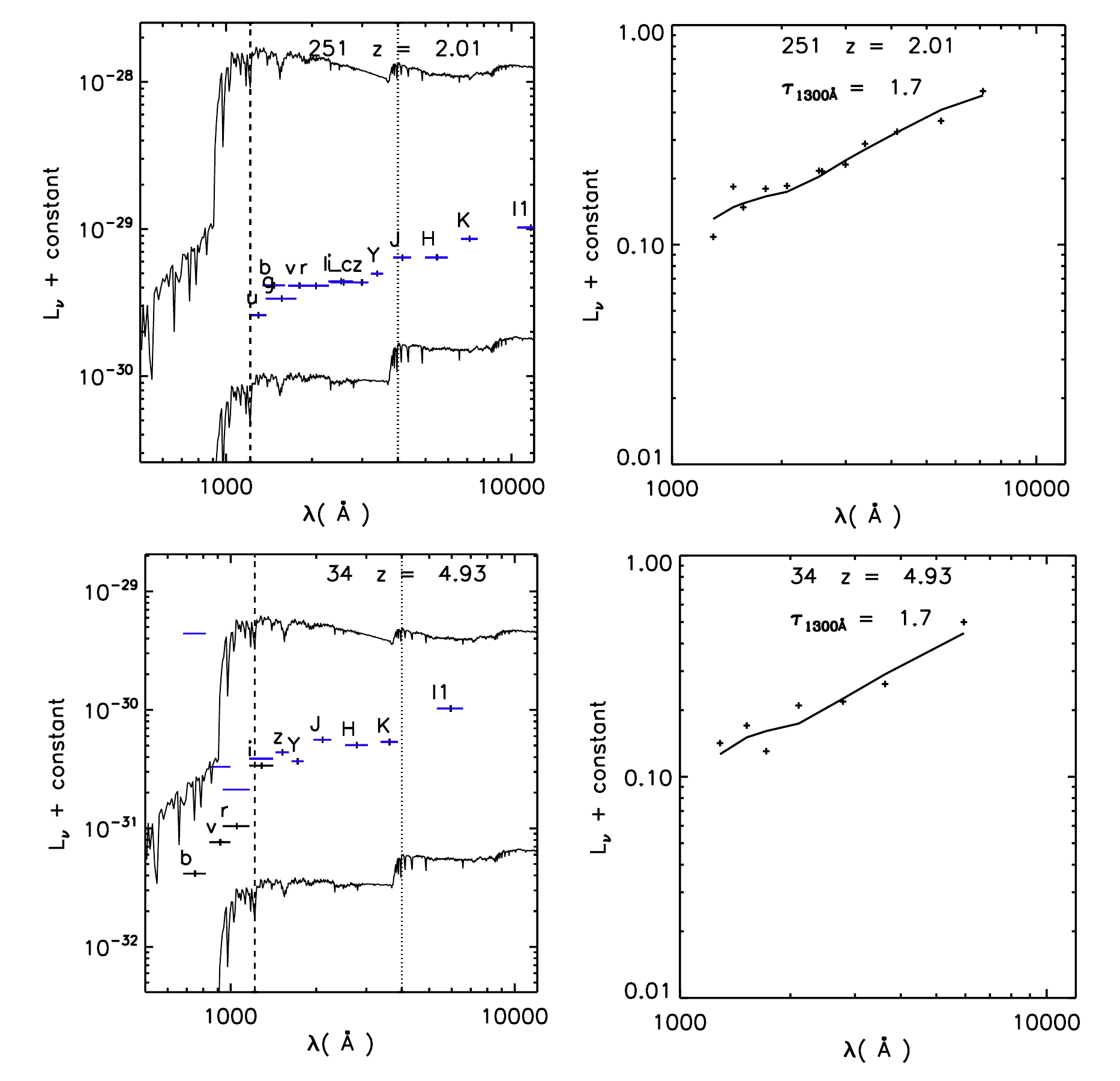}
\caption{Left -- The rest frame SEDs are shown for two sources at z = 2.01 and 4.93. The constant rate SB spectra are
shown for ages of 50 and 500 Myr for comparison. The dash and dotted lines correspond to 1216 \AA~and 4000 \AA. Right -- The best-fit SEDs using the derived
wavelength dependence of the attenuation (normalized to the reference wavelength $\lambda = 1300$ \AA) and an absolute value of the optical depth $\tau_{1300\AA} = 1.7$ (coincidentally in both instances). Although the best-fit curve extends to 950 \AA, these plots
limit the spectral range of the curve to wavelengths for which each source has photometry.  \label{sample_source}}
\end{figure*}

The COSMOS photometry used here includes the 34 band photometry used for the COSMOS photometric redshifts \citep{ilb13}
and the near infrared deep photometry from UltraVista \citep{mcc12} and Spitzer SPLASH \citep{ste14}. For the attenuation 
curve analysis here, we use 14 broad bands (u$_{CFHT}$, B$_{Subaru}$, V$_{Subaru}$, g$_{Subaru}$, r$_{Subaru}$, i$_{Subaru}$, z$_{Subaru}$, I$_{CFHT}$, IRAC1, IRAC2, Y$_{Ultra-Vista}$, J$_{Ultra-Vista}$, H$_{Ultra-Vista}$, K$_{Ultra-Vista}$) which have the highest signal-to-noise ratios \citep[see Table 1 - ][]{ilb09}. At least 6 bands were required 
in the rest wavelength range between Lyman $\alpha$ (1216 \AA) and 10000 \AA. In addition, we required at least one detection 
on each side of the Balmer/4000 \AA~break. 

For rest wavelengths short of Lyman $\alpha$ there is significant intergalactic medium HI absorption at the higher redshifts. At these wavelengths we corrected the photometry to brighter values using the standard Madau correction \citep{mad95}. This correction has a very large dispersion \citep[see][]{mad95} and we found that it clearly overcorrected
the photometry, as indicated from the fact that the corrected fluxes short of Ly $\alpha$ exceeded those of adjacent bands just longward of Ly $\alpha$. Reducing the Madau correction by 50\% avoids this. The derived dust attenuation 
at the one point short of 1216 \AA~should therefore be viewed with some uncertainty since it hinges on the values of the IGM HI correction. 

Sample SEDs for two such objects at z = 2.01 and 4.93 are shown on the left in Figure \ref{sample_source} together 
with the constant rate starburst spectra for 50 and 500 Myr duration for reference.

\begin{figure}[ht]
\epsscale{1.}
\plotone{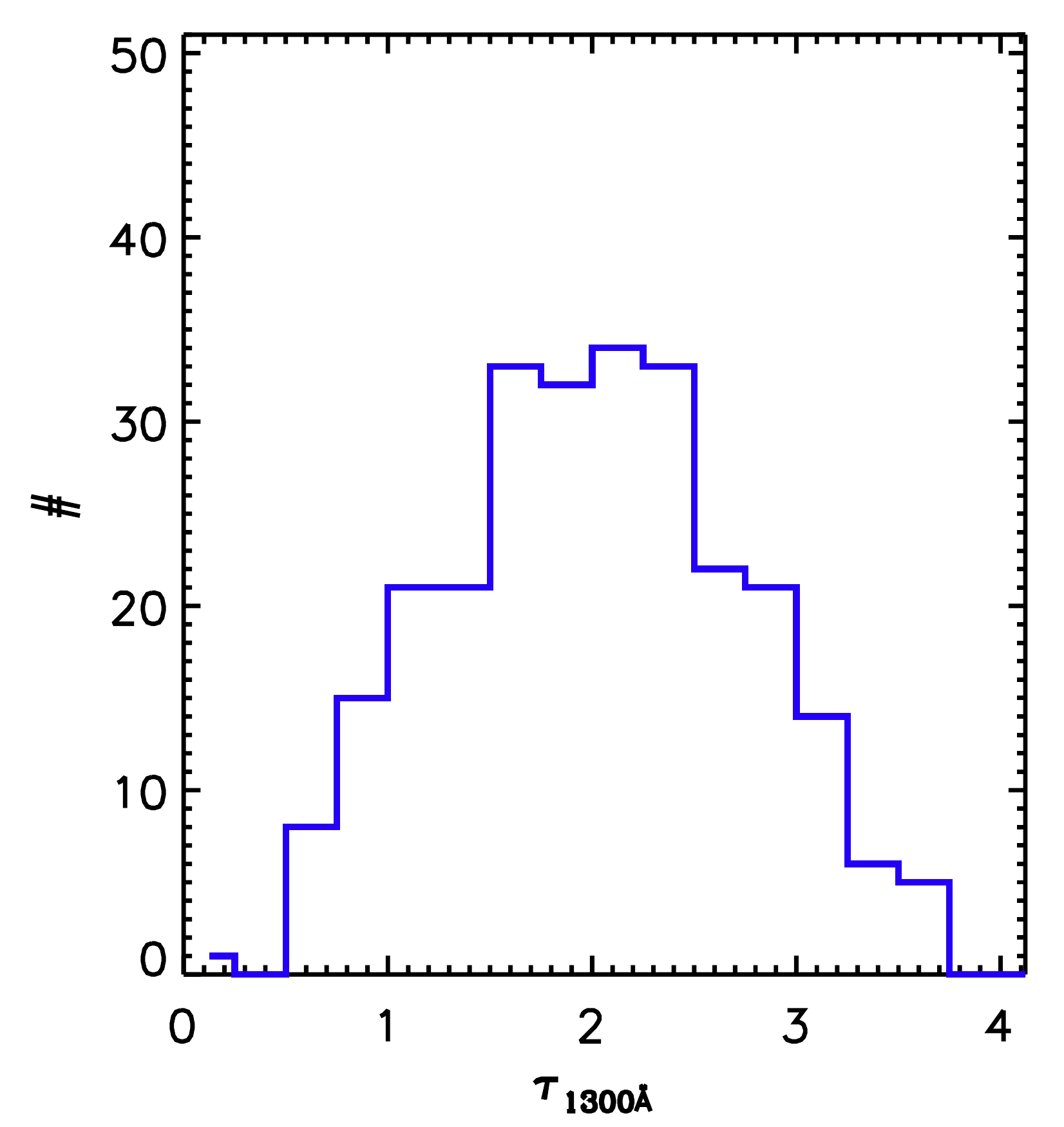}
\caption{ The distribution of derived $\tau_{1300\AA}$ is shown for the full sample at z = 2 to 6.5.   \label{taus}}
\end{figure}

\section{Numerical Solution}\label{technique}

We assume that the dust attenuation has a wavelength dependence common to all sources in each redshift sample, but that the absolute
value of the dust optical depth at a fiducial wavelength is freely varying between sources. Thus, the wavelength dependence is characterized by a 
step function ($\tau (\lambda_i)$) with discrete wavelength bins spanning rest frame $\lambda = $950 and 10000 \AA. This form of fitting has the virtue 
of making no assumption or analytic parametrization of the extinction. In addition, there is an absolute source-dependent scale factor for the optical depth at rest frame $\lambda = 1300 \AA~(\tau_{1300\AA}$). Thus, 

\begin{eqnarray}
\tau_{\lambda} (\rm source) &=& \rm \tau_{1300\AA} (\rm source)~ \tau (\lambda_i)  .  
 \end{eqnarray}
 
\noindent  For the normalized opacity function ($\tau (\lambda_i)$), we use 9 bins spaced logarithmically in wavelength. If the rest-frame specific luminosity of the unobscured starburst
is denoted as $l_{\nu} (\rm SB)$, then the model we fit to the observed SEDs is given by 
 
\begin{eqnarray}
s_{\nu} = e^{-\tau_{\lambda} (\rm source)} ~\rm L _0 (\rm source)~ l_{\nu} (SB) . 
 \end{eqnarray}
 \noindent The underlying source SED ($l_{\nu} (SB)$) is taken to be the 50 Myr starburst spectrum shown in Figure \ref{spectra} based on our deselection of sources with larger Balmer Break, as measured from the best-fit stellar population model to the photometry.  We did experiment with using the 20 and 100 Myr SED and they did not improve the resultant best fit $\chi^2$. 
 
For the z = 2 to 6.5 sample of 266 galaxies, the number of free parameters to be solved for in minimized $\chi^2$ is: 
266 ($L_0 (\rm source) $) + 266 ($\tau_{1300\AA} (\rm source))$ + 9 (wavelength bins) = 541 parameters. For this minimization we used the
 Levenberg-Marquardt least-squares minimization routine (MPFIT in IDL). We adopted uncertainties of 10\% for each SED data point; this was done to give all SED points similar weight in the fitting. Thus, we avoid  
 overweighting a single band which has high signal-to-noise ratio but which does not provide much wavelength leverage. This 
 ensures that the IRAC data (3.6$\mu$m  and 5.8$\mu$m) constrain the fit at long wavelengths.
 
  The least-squares fitting was done in two passes: first using the complete sample of galaxies and then removing the 
 few objects ($\sim10$) which had T$_{1300} > 5$; these latter sources are likely to have low signal-to-noise ratio at the short wavelengths. The derived results  were not significantly different after the second pass except that the $\chi^2$ was improved slightly ($\leq 10$\%). Figure \ref{taus} shows the distribution of derived $\tau_{1300\AA}$ of the final solution.

\begin{figure*}[ht]\label{full_extinct_plot}
\epsscale{0.7}
\plotone{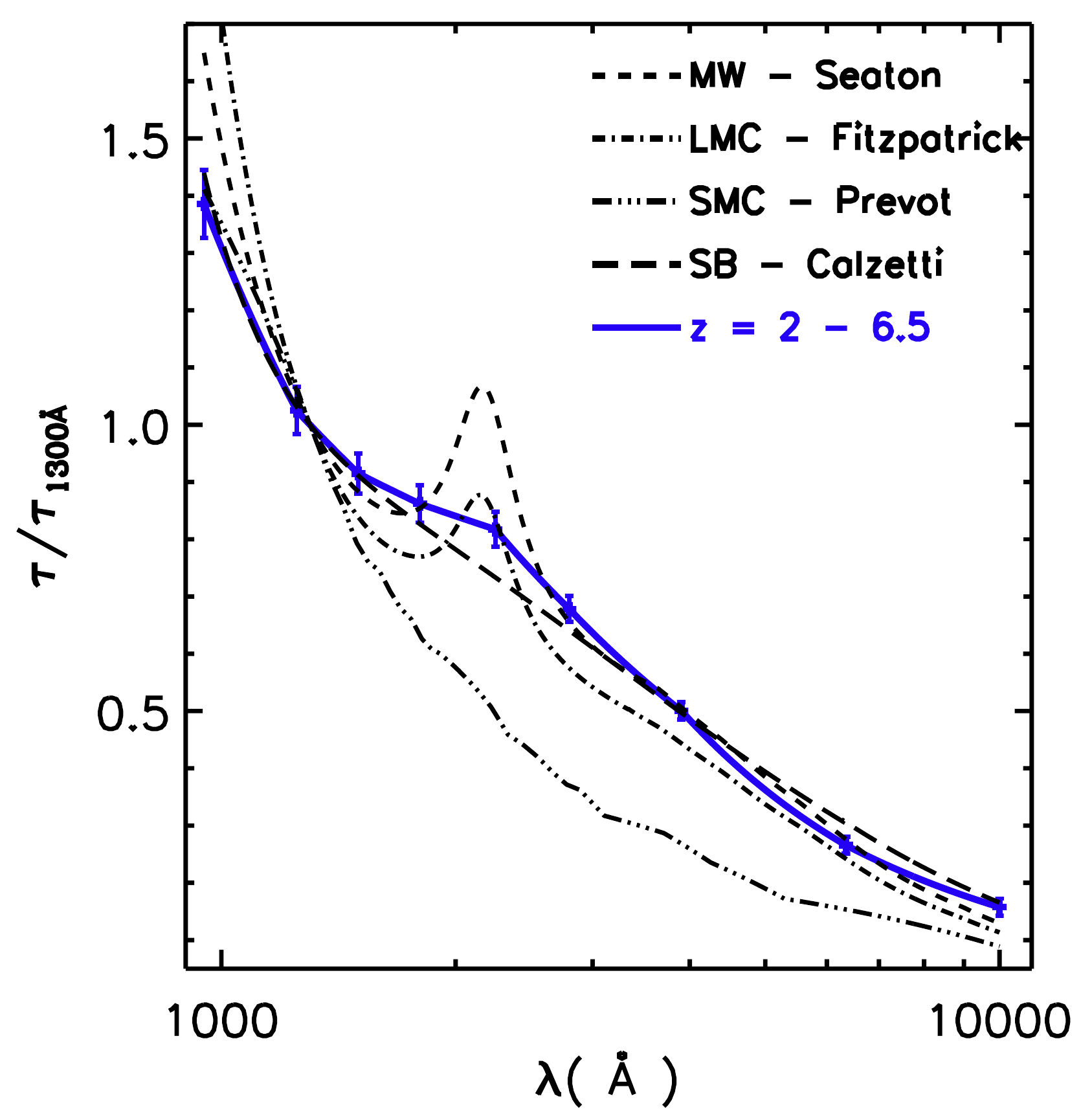}
\caption{The wavelength dependence of the attenuation is shown for the full sample spanning z = 2 to 6.5 with uncertainties determined 
from the Levenberg-Marquardt routine. Also shown are the 
extinction curves for the Milky Way \citep{sea79}, LMC \citep{fit86}, SMC \citep{pre84} and local starburst galaxies \citep{cal00}. The wavelength dependence derived here for galaxies at z $ > 2$ is similar to that of Calzetti curve  derived for starburst galaxies but also includes the 2175 \AA~bump feature as seen in the Milky Way and LMC extinction curves.\label{extinct}}
\end{figure*}

\begin{figure*}[ht]
\epsscale{1.}
\plotone{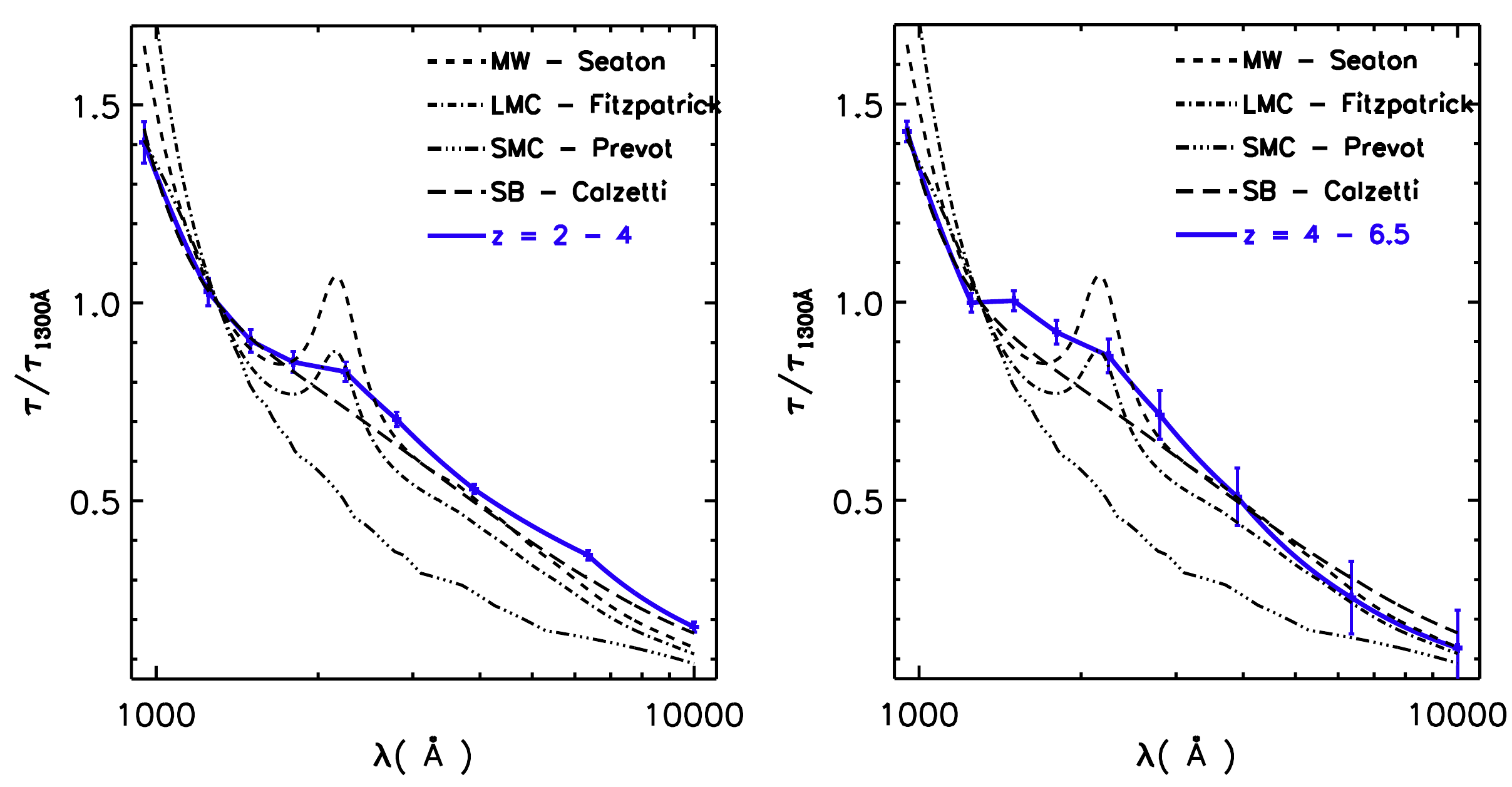}
\caption{The wavelength dependence of the attenuation is shown for sub-samples at z = 2 - 4 (Left) and 4 - 6.5 (Right). 
Given the larger uncertainties in the z  = 4 - 6.5 fit, we do not think the difference in the 2175 \AA~bump feature is significant. The overall 
broadband wavelength dependencies are more consistent with the MW, LMC and SB curves and not with the SMC curve.  \label{extinct_z}}
\end{figure*}

\begin{deluxetable}{cccccccc}
\tabletypesize{}
\tablecaption{Best-fit Attenuation Curve \label{curve}  }
\tablewidth{0pt}
\tablehead{\colhead{$\lambda$} & \colhead{z = 2 - 6.5} & \colhead{z = 2 - 4} & \colhead{z = 4 - 6.5}     \\
\colhead{\AA}  & \colhead{$\tau_{\lambda}/\tau_{1300}$}  & \colhead{$\tau_{\lambda}/\tau_{1300}$} &  \colhead{$\tau_{\lambda}/\tau_{1300}$}}
\startdata
\\
950   &  1.386$\pm$0.060  &      1.406$\pm$0.053   &  1.431$\pm$0.026 \\   
1250     &  1.025$\pm$0.041  &     1.028$\pm$0.035    &  0.999$\pm$0.024  \\       
1500   &  0.915$\pm$0.035  &     0.904$\pm$0.029    &   1.003$\pm$0.025 \\         
1800   &  0.862$\pm$0.033  &      0.850$\pm$0.026   &   0.924$\pm$0.030 \\         
2250   &  0.817$\pm$0.030 &      0.826$\pm$0.025    &  0.865$\pm$0.043 \\   
 2800     & 0.678$\pm$0.023   &     0.706$\pm$0.019    &  0.716$\pm$0.061  \\        
 3900     &  0.501$\pm$0.015  &     0.530$\pm$0.012    &   0.509$\pm$0.073 \\        
 6350       &  0.266$\pm$0.016  &      0.363$\pm$0.011   &   0.255$\pm$0.092 \\     
 10000   &  0.158$\pm$0.017  &       0.181$\pm$0.014  &    0.127$\pm$0.096 \\ 
 \\
 \hline
 \\
\# of galaxies\tablenotemark{a} & 266 & 135  & 132 \\ 
\# of phot. bands\tablenotemark{b} & 2747 & 1614 & 1133 \\
$\chi^2$ & 3.68 & 1.92 & 6.63  \\
\enddata
\tablecomments{As described in the text, the least-squares fitting solved for the wavelength 
dependence as a step function so that each wavelength point is independent, and smoothness of continuity 
is not enforced except if warranted by the data. The attenuation curves were each normalized to unity at $\lambda = 1330$ \AA. }
\tablenotetext{ a}{Number of galaxies in the final samples for the fit. The samples were slightly larger 
on the first pass fitting; after the first pass 5-15 galaxies with $\tau_{1300\AA} >5$  were removed for the second pass fitting since their short wavelength points would have low signal-to-noise ratio. Removing these sources
reduced the $\chi^2$ but did not significantly change the shape of the attenuation  curve. }
\tablenotetext{ a}{Total number of photometric band fluxes used in fitting.}
\end{deluxetable}

 \section{Results}\label{results}
 
 The derived spectral dependence of the normalized dust attenuation is shown in Figure \ref{full_extinct_plot} for the full 
 sample at z = 2 to 6.5. The uncertainties on the attenuation curve shown in the figure are from the Levenberg-Marquardt algorithm. 
 The $\chi^2$ for the fits in the three redshift ranges are given in Table \ref{curve} are 3.7, 1.9 and 6.6. For the fitting, the uncertainties 
 on each flux measurement were set to 10\%, thus the derived fits typically disperse only $13 - 25$\% from the observed fluxes on all of the individual sources. 
 
 Also shown are the extinction curves for the Milky Way, LMC and SMC plus the dust attenuation 
 curve for starburst galaxies from \cite{cal00}. All curves were normalized to the their values at 1300 \AA. The overall wavelength 
 dependence of the attenuation 
 curve for z = 2 to 6.5 is remarkably similar to that of the low redshift attenuation curve for the SB galaxies. It is also very similar to  
 the extinction curves for the Milky Way and LMC, but not the SMC which has a much stronger wavelength dependence in the UV.  Table \ref{curve} 
 lists the numeric values for the derived attenuation curves. 

 The high redshift curve (Figure \ref{full_extinct_plot}, Table \ref{curve}) also clearly shows the presence of the 2175 \AA~bump feature seen in the Milky Way and the LMC 
 extinction curves but not in the starbursts or the SMC. This feature is often attributed to graphite (hence diamonds in our title) grains and its absence in the SMC extinction curve is probably related to the low metallicity of SMC, resulting in different grain properties. The 
 absence of the 2175 \AA ~bump feature in the Calzetti curve (derived for local starburst galaxies) is certainly significant in \cite{cal94} (see Figure 17 there). 
 
 In order to probe the redshift evolution of the attenuation, we also obtained solutions for sub-samples of the galaxies 
 at z = 2 - 4 and z = 4 - 6.5. The results are shown in Figure \ref{extinct_z}. Their attenuation curves do not appear significantly 
 different from that shown in Figure \ref{extinct} for the full sample. The z = 4 - 6.5 curve exhibits a somewhat larger width 
 for the 2175 \AA~bump feature but we don't consider this significant given the uncertainties.

\section{Summary and Comments} \label{discussion} 

Our technique uses galaxies selected spectroscopically with strong CIV absorption and photometrically with a small Balmer/4000 \AA~ break feature to yield a sample of galaxies for which the un-extincted UV-optical SEDs are dominated by OB stars. These SEDs 
have a nearly constant shape since their emission over the entire wavelength is from a single class of stars. 
Our sample of galaxies taken from the large COSMOS survey field; it consists of 266 sources after removing sources with 
incomplete wavelength coverage. 

The derived attenuation is very similar to the \cite{cal00} curve obtained for 40 low redshift starburst galaxies. The 
curve shown in Figure \ref{full_extinct_plot} clearly shows the 2175 \AA~bump feature; this was not present in the Calzetti curve, but is seen in both the Milky Way 
and LMC. Our curve is tabulated in Table \ref{curve} for use in correcting colors and extinctions from high redshift galaxy surveys -- 
both star forming and non-star forming galaxies. 

It is important to remember that the curves shown here for MW, LMC and SMC were all obtained 
 from line-of-sight extinction measurements of stars. In contrast, the starburst curve \citep{cal00} and that derived here for high redshift
 are 'effective attenuation' curves -- they are obtained from the integrated galaxy light. The former extinction curves include the 
 effects of both dust absorption and scattering while the effective attenuation curves may contain a lesser contribution from dust 
 scattering (since some photons are scattered back into our line-of-sight). The attenuation curves also include the very important 
 effect of partial covering of the galactic continuum by dust in the line-of-sight. This is of course not 
 relevant to extinction measurements of individual stars. 
 
 Two other instances where the high redshift attenuation/extinction curve has been analyzed are for GRB sources and QSOs. 
 \cite{eli09} obtained an extinction curve for GRB 070802 at z = 2.45 which was similar to the LMC extinction curve shown in 
 Figure \ref{full_extinct_plot}, including the 2175 \AA~bump at high signal-to-noise ratio. \cite{per11} also clearly detected the 2175 \AA~
 feature in GRB 080607 at z =3.03.  
 
 In contrast to the GRBs and the starburst galaxies analyzed here, the high redshift QSOs show 
 quite different extinction curves.  They generally resemble the SMC extinction curve \citep{ric03,rei03,hop04a}, not the LMC, Milky Way or starburst curves. This said, \cite{gal10} analyzed the dust extinction in a sample of 39 quasars at z  = 3.9 to 6.4 and found a substantially flatter extinction curve at 
 $\lambda < 1300$ \AA~than the SMC curve. In fact, their extinction curve is flatter than the Calzetti curve at $\lambda < 2500$ \AA, implying that it is also flatter than that obtained here at high redshift. They see no evidence of the 2175 \AA~bump feature, although in many of their quasars this feature might be difficult to detect due to the broad QSO emission lines. In the quasars, the obscuration occurs from
 three distinct regions: the inner region with gas only seen in the X-rays, the mid-IR very hot dust within a few parsec of the AGN and
 the host galaxy dust \citep{elv12}. In the first two regions, the dust abundance and composition is likely to be very different. 

 An alternative, future application of the technique developed here would be possible if high quality spectra became available over the UV-opt wavelength range 
on a small number of bright sources. Then, accurate measures of the equivalent width of the CIV spectral feature and the Balmer break strength could be used 
select more precisely the source SED templates for each individual source.

\acknowledgments
We thank the referee for suggestions and Martin Elvis for comments on an early draft. We also thank 
 Zara Scoville for proof reading the manuscript. A.F. acknowledges support, from the Swiss National Science Foundation and thanks Caltech for hospitality while working on this article.


\bibliography{scoville_extinct}{}





\end{document}